\documentclass{mn2e}
\usepackage{psfig}

\title[Hierarchical galaxy clustering in the 2dFGRS]
{\vspace{-0.5cm}The 2dF Galaxy Redshift Survey: Hierarchical galaxy clustering
\vspace{-0.5cm}
}
\author[Baugh et al.]{
\parbox[t]{\textwidth}{
C. M. Baugh$^1$,
D. J. Croton$^{2}$,
E. Gazta\~{n}aga$^{3,4}$,
P. Norberg$^{5}$,
M. Colless$^6$,
I. K.\ Baldry$^7$,
J. Bland-Hawthorn$^6$,
T. Bridges$^{8,6}$, 
R. Cannon$^6$, 
S. Cole$^1$, 
C. Collins$^{9}$, 
W. Couch$^{10}$, 
G. Dalton$^{11,12}$,
R. De Propris$^{13}$,
S. P.\ Driver$^{13}$, 
G. Efstathiou$^{14}$, 
R. S.\ Ellis$^{15}$, 
C. S.\ Frenk$^1$, 
K. Glazebrook$^7$, 
C. Jackson$^{16}$,
O. Lahav$^{17,14}$, 
I. Lewis$^{11}$, 
S. Lumsden$^{18}$, 
S. Maddox$^{19}$,
D. Madgwick$^{20}$,
J. A.\ Peacock$^{21}$,
B. A.\ Peterson$^6$, 
W. Sutherland$^{14}$,
K. Taylor$^{15}$.
(The 2dFGRS Team)
}
\vspace*{6pt} \\ 
$^1$Department of Physics, University of Durham, South Road, 
    Durham DH1 3LE, UK. \\ 
$^2$Max-Planck-Institut f\"ur Astrophysik, D-85740 Garching, Germany. \\
$^3$INAOE, Astrofisica, Tonantzintla, Apdo Postal 216 y 51, Puebla 7200,
    Mexico. \\
$^4$Institut d'Estudis Espacials de Catalunya, ICE/CSIC, Edf.
    Nexus-104-c/Gran Capita 2-4, 08034 Barcelona, Spain.  \\
$^5$ETHZ Institut f\"ur Astronomie, HPF G3.1, ETH H\"onggerberg, CH-8093
       Z\"urich, Switzerland. \\
$^6$Anglo-Australian Observatory, P.O.\ Box 296, Epping, NSW 2111,
    Australia.\\  
$^7$Department of Physics \& Astronomy, Johns Hopkins University,
       Baltimore, MD 21118-2686, USA. \\
$^{8}$Department of Physics, Queen's University, Kingston, 
    Ontario K7L 3N6, Canada. \\
$^{9}$Astrophysics Research Institute, Liverpool John Moores University,  
    Twelve Quays House, Birkenhead, L14 1LD, UK. \\
$^{10}$Department of Astrophysics, University of New South Wales, Sydney, 
    NSW 2052, Australia. \\
$^{11}$Department of Physics, University of Oxford, Keble Road, 
    Oxford OX1 3RH, UK. \\
$^{12}$Space Science \& Technology Division, Rutherford Appleton Laboratory, 
    Chilton OX11 0QX, UK. \\
$^{13}$Research School of Astronomy \& Astrophysics, The Australian 
    National University, Weston Creek, ACT 2611, Australia. \\
$^{14}$Institute of Astronomy, University of Cambridge, Madingley Road,
    Cambridge CB3 0HA, UK. \\
$^{15}$Department of Astronomy, California Institute of Technology, 
    Pasadena, CA 91025, USA. \\
$^{16}$CSIRO Australia Telescope National Facility, PO
    Box 76, Epping, NSW 1710, Australia. \\
$^{17}$Department of Physics and Astronomy, University College London, 
    Gower Street, London WC1E 6BT, UK. \\
$^{18}$Department of Physics, University of Leeds, Woodhouse Lane,
       Leeds, LS2 9JT, UK. \\
$^{19}$School of Physics \& Astronomy, University of Nottingham,
       Nottingham NG7 2RD, UK. \\
$^{20}$Department of Astronomy, University of California, Berkeley, 
       CA 94720, USA. \\
$^{21}$Institute for Astronomy, University of Edinburgh, Royal Observatory, 
       Blackford Hill, Edinburgh EH9 3HJ, UK.
\vspace{-0.5cm}
}

\newcommand{\plotone}[1]
           {\centering \leavevmode \psfig{file=#1,width=\columnwidth,clip=}}

\begin{document}

\maketitle

\begin{abstract}
We use the two-degree field Galaxy Redshift Survey (2dFGRS) to test
the hierarchical scaling hypothesis: namely, that the $p$-point 
galaxy correlation functions can be written in terms of the two point 
correlation function  or variance. 
This scaling is expected if an initially
Gaussian distribution of density fluctuations evolves under the action
of gravitational instability. We measure the volume averaged $p$-point
correlation functions using a counts in cells technique applied to a
volume limited sample of 44,931 $L_*$ galaxies.  We demonstrate that
$L_{*}$ galaxies display hierarchical clustering up to order $p=6$ in
redshift space.  
The variance measured for $L_{*}$ galaxies is in excellent agreement 
with the predictions from a $\Lambda$-cold dark matter N-body simulation.
This applies to all cell radii considered, 
$0.3<(R/h^{-1}{\rm Mpc})<30$.
However, the higher order correlation functions of $L_*$ galaxies have 
a significantly smaller amplitude than is predicted for the dark 
matter for $R<10h^{-1}$Mpc. 
This disagreement implies that a non-linear bias exists between the 
dark matter and $L_*$ galaxies on these scales.
We also show that the presence of two rare, massive
superclusters in the 2dFGRS has an impact on the higher-order clustering 
moments measured on large scales.
\end{abstract}

\begin{keywords}
cosmology: observations, large-scale structure of Universe
\vspace{-0.7cm}
\end{keywords}

\section{Introduction}

Current theoretical models of structure formation in the Universe are
based on the paradigm of gravitational instability. This process is
believed to be responsible for driving the growth of small primordial
density perturbations into the nonlinear collapsed structures such as
galaxies and clusters that are evident in the Universe today.

The premise of gravitational instability has been tested
indirectly by comparing the clustering predicted by numerical
simulations of the formation of cosmic structures with the observed
distribution of galaxies (e.g. Benson et al. 2001). A direct test of
this fundamental ingredient of structure formation models was made
using the 2dFGRS by Peacock et al.  (2001). The size of the 2dFGRS
allowed the first accurate measurement of the
two-point galaxy correlation function on large scales. Peacock et
al. demonstrated that the two-point correlation function at large pair
separations displays a form that is characteristic of the bulk motions
of galaxies expected in the gravitational instability scenario.

We present an independent test of the gravitational
instability paradigm. For a Gaussian distribution of
density fluctuations, the volume averaged correlation functions,
$\bar{\xi}_{p}$, are identically zero for $p>2$; the density field
is completely described by its variance, $\bar{\xi}_{2}$. The
evolution of an initially Gaussian density field due to gravitational
instability generates non-zero $\bar{\xi}_{p}$
(Peebles 1980).  A basic test of the gravitational origin of the
higher order moments is to determine their relation to the variance of
the distribution.  This is traditionally encapsulated in the
hierarchical model:
\begin{equation}
\bar{\xi}_{p} = S_{p} \bar{\xi}_{2}^{p-1}.
\label{eq:sp}
\end{equation}
This model applies to real space clustering; however, in redshift space the
scaling still tends to hold even on small scales where
the ``fingers-of-God'' effect is prominent (Lahav et al. 1993; Hoyle, 
Szapudi \& Baugh 2000).
Perturbation theory predicts that the hierarchical amplitudes 
for the mass distribution are independent 
of the cosmological density parameter, the cosmological constant 
and cosmic epoch (Bernardeau et al. 2002).  

We use the 2dFGRS (Colless et al. 2001, 2003) to measure the higher 
order correlation functions of the galaxy distribution, focusing on the
clustering of $L_*$ galaxies.  The size of the 2dFGRS
is exploited to extract a volume limited sample of $L_*$ galaxies,
which greatly simplifies our analysis (Section 2). The results for the
volume averaged correlation functions, up to sixth order, are
presented in Section 3, in which we also test how well the
hierarchical scaling model works.  Our conclusions are given in
Section 4.

\section{Data and analysis}

The density of galaxies is a strong function of radial distance in a
magnitude limited survey. This needs to be compensated for in
any clustering analysis by applying a suitable weighting scheme
(e.g. Saunders et al. 1991). Alternatively, one may construct a volume
limited sample by selecting certain galaxies from the full
redshift survey.  These galaxies are chosen so that they would appear
inside the apparent magnitude range of the survey if displaced to any
redshift within the interval defining the sample. The only radial
variation in galaxy number density in a volume limited sample is due
to large scale structure in the galaxy distribution.  This makes
volume limited samples much more straightforward to analyse than flux
limited samples. However, only a fraction of the galaxies from the 
full redshift survey satisfy the selection criteria in redshift and 
absolute magnitude. This reduction in the density of
galaxies has curtailed the utility of volume limited subsamples
constructed from earlier redshift surveys.

We construct a volume limited sample of $L_*$ galaxies
from the 2dFGRS. The motivation for the choice of a sample centred on
$L_*$ is clear; this results in a volume limited sample
with the largest possible number of galaxies for magnitude bins of a
given size. As the luminosity used to define a sample increases, the
selected galaxies can be seen out to larger redshifts and thus sample
larger volumes. However, brighter than $L_*$, the space density of
galaxies drops exponentially (e.g. Norberg et al. 2002). Hence, the
optimum balance between volume surveyed and intrinsic galaxy space
density is achieved for $L_*$ galaxies.  In addition, the higher order
clustering of $L_*$ galaxies provides a benchmark or reference
against which to compare trends in clustering strength with galaxy
luminosity (see Norberg et al 2001; Croton et al. 2004a).  We consider
the two contiguous areas of the 2dFGRS, referred to as the NGP and SGP
regions, which contain around 190,000 galaxies with redshifts and
cover an effective area of approximately 1200 square degrees in
total. After selecting galaxies with absolute magnitudes in the range
$-19>M_{b_{\rm J}}-5\log_{10}h>-20$ (corrected to $z=0$ using the
global $k+e$ correction quoted by Norberg et al. 2002), the volume
limited sample contains 44,931 galaxies.  The redshift interval of the
sample is $z=0.021$ to $0.130$, corresponding to a volume of
$7.97 \times 10^{6}h^{-3}{\rm Mpc}^{3}$ for the combined NGP and SGP
regions.

\subsection{Counts in cells analysis}

The distribution of counts in cells is estimated by throwing down a
large number of spherical cells, on the order of $10^7$ for each
cell radius considered, within the $L_*$ volume limited 2dFGRS sample.
Full details of how we deal with the spectroscopic incompleteness and
the angular mask are given in Croton et al. (2004a); the corrections
turn out to be small in any case (see figure 1 of Croton et al.).

The higher order correlation functions, $\bar{\xi}_{p}$, are the
reduced $p^{\rm th}$ order moments of the distribution of galaxy counts in
cells.  The estimation of the higher order correlation functions from
the cell count probability distribution is explained in a number of
papers (e.g. Gazta\~{n}aga 1994; Baugh, Gazta\~{n}aga \& Efstathiou
1995; Croton et al. 2004a).  The variance or width of the count
distribution is given by the case $p=2$.  For $p>2$, 
the correlation functions probe further out into the tail
of the count probability distribution.

We use mock 2dFGRS catalogues to estimate the errors
on the measured higher order correlation functions. Full details
of the mocks can be found in Norberg et
al. (2002) and Croton et al. (2004a).

\section{Results}

\begin{figure*}
\begin{picture}(250,280)
\put(-100,0){\psfig{file=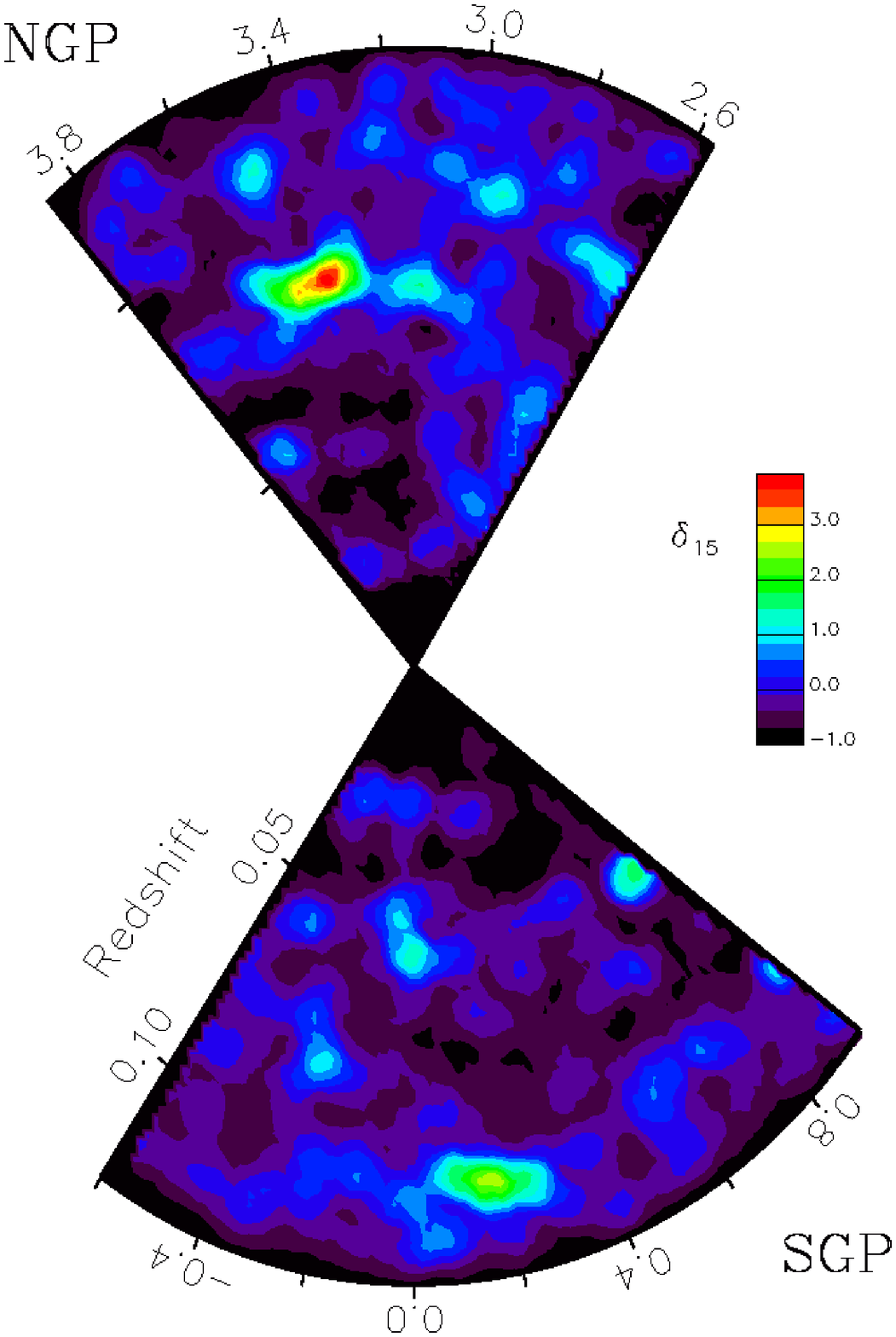,width=7cm,height=10.1cm}}
\put(145,0){\psfig{file=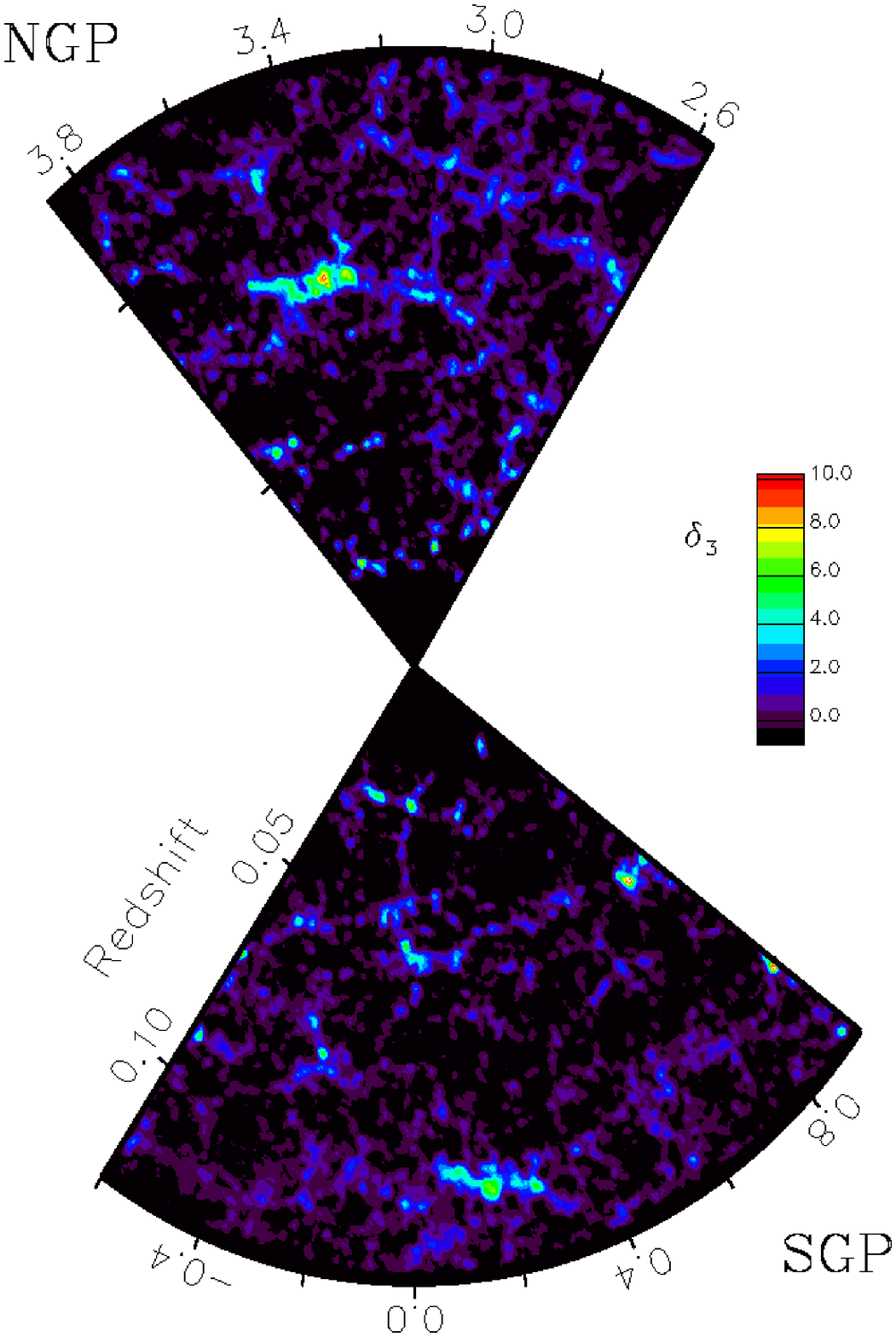,width=7cm,height=10.1cm}}
\end{picture}
\caption{ 
  The projected galaxy density in the $L_*$ volume limited sample,
  smoothed on two different scales.  The galaxy distribution is
  projected onto the right ascension--redshift plane and is then
  smoothed in circular cells of radius $15h^{-1}$Mpc (left)
  and $3h^{-1}$Mpc radius (right). Redder
  colours denote higher densities as indicated by the key that
  accompanies each panel.  Two ``hot-spots'' stand out clearly,
  particularly in the left hand coneplots; one in the NGP at $z\approx
  0.08$ and the other in the SGP at $z \approx 0.11$.  
  Right ascension is given in radians.
  }
\label{fig:cone}
\end{figure*}

The projected density of galaxies in the $L_*$ volume limited sample
is shown in Fig.~\ref{fig:cone}. The galaxy density projected onto the
right ascension--redshift plane is smoothed using circular windows. 
Two different smoothing radii have been used to produce these maps;
the left-hand panel shows the density after smoothing with a
circular cell of radius $15h^{-1}$Mpc and the right hand panel shows
the distribution as sampled with a cell of radius $3h^{-1}$Mpc. The
redder colours indicate higher galaxy densities, as shown by the
scale that accompanies each cone plot.  Two `hot-spots' are readily
apparent, particularly in the cone plot smoothed on the larger
scale. These correspond to superclusters of galaxies that also appear
in the 2dFGRS Percolation Inferred Galaxy Group catalogue (Eke et al.
2004) and in the reconstructed density field of the 2dFGRS (Erdogdu et
al. 2004).  The presence of these superclusters has an impact on the
extreme event tail of the count probability distribution. Later in
this section, we will investigate the influence of these structures on
our measurement of the higher order correlation functions by excising
the volumes that contain the superclusters from our analysis. The
`cosmic web' of filamentary structures and voids is apparent in the
cone smoothed on the smaller scale.

\begin{figure}
\plotone{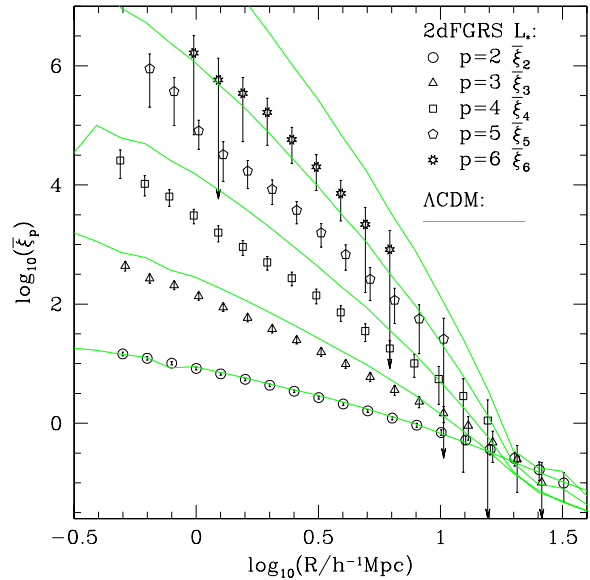}
\caption{ 
  The higher order correlation functions $\bar{\xi}_{p}$ measured for
  $L_{*}$ galaxies in the 2dFGRS (symbols).  The orders $p=2$--$6$ are
  shown, as indicated by the key.  The errorbars show the {\it rms}
  scatter estimated using mock 2dFGRS catalogues.  The lines show the
  $\bar{\xi}_{p}$ measured for the dark matter in redshift space in
  the $\Lambda$CDM Hubble Volume simulation, for orders $p=2$ to $6$
  in sequence of increasing amplitude for $R<10h^{-1}$Mpc.
  }
\label{fig:xip1}
\end{figure}

The higher order correlation functions measured for $L_*$ galaxies are
plotted in Fig.~\ref{fig:xip1}. The correlation functions are only
plotted on scales for which a robust measurement is possible.  The
correlation functions show a dramatic steepening on small scales as
the order $p$ increases. For example, the ratio 
$\bar{\xi}_{6}/\bar{\xi}_{2}$ is $10^5$ at $R=1h^{-1}$Mpc,
falling to $\sim 100$ at $R=6.3h^{-1}$Mpc.  We also plot
the higher order correlation functions for the dark matter
distribution in the $\Lambda$CDM Hubble Volume simulation (Evrard et
al. 2002).  These theoretical predictions include the effects of
peculiar motions in the distant observer approximation.  The variance
of the dark matter in redshift space agrees spectacularly well with
the measured $\bar{\xi}_{2}$ for $L_{*}$ galaxies.  This confirms 
the conclusions reached in independent analyses of the clustering 
of $L_*$ galaxies in the 2dFGRS (Lahav et al. 2002; Verde et al. 2002).  
However, for the case of the
$\Lambda$CDM Hubble Volume simulation, the $p>2$ moments of the dark
matter differ from the measurements for $L_*$ galaxies for $R<10h^{-1}$Mpc.

\begin{figure}
\plotone{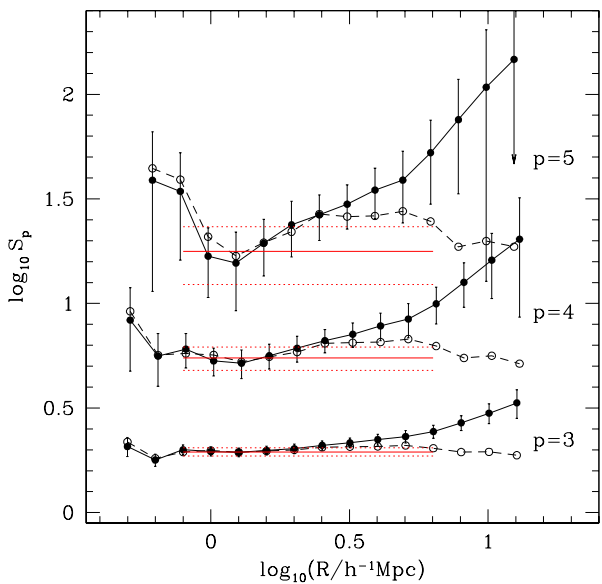}
\caption{
  The hierarchical amplitudes, $S_p$, for $p=3,4$ and $5$, plotted as
  a function of cell radius for the $L_*$ volume limited sample.  The
  filled symbols connected by solid lines show the results obtained
  using the full volume.  The best fit constant values of $S_p$ are
  shown by the horizontal solid lines, which are 
  plotted over the range of scales
  used in the fit.  The dotted lines show the 1$-\sigma$ error on the fit.  The
  open symbols connected by dashed lines show the hierarchical
  amplitudes recovered when the two largest superclusters are masked
  out of the volume.
  }
\label{fig:sp2}
\end{figure}

The hierarchical amplitudes, $S_p$, obtained from the $\bar{\xi}_{p}$
by applying Eq. \ref{eq:sp} are plotted as a function of cell radius
for orders $p=3$--$5$ in Fig.~\ref{fig:sp2} ($p=6$ is omitted for
clarity).  For $p=3$, $S_3$ is approximately constant for
cells with $R<3h^{-1}$Mpc. At larger $R$, $S_3$ increases with
radius. This behaviour is mirrored for $p>3$, with the upturn in $S_p$
seen at progressively smaller radii as $p$ increases.
Perturbation theory predicts that, on large scales, the $S_p$ should
have only a weak dependence on scale for CDM-like power spectra
(Juszkiewicz, Bouchet \& Colombi 1993).  In redshift space, the
hierarchical amplitudes are expected to be approximately independent
of scale over an even wider range of scales than those on which
perturbation theory is applicable (Hoyle et al. 2000;
Bernardeau et al. 2004).  We therefore attempt to fit a constant value
of $S_p$ to the ratios plotted in Fig.~\ref{fig:sp2}. We use a
principal component analysis to take into account the correlation 
between the $\bar{\xi}_{p}$ in neighbouring bins (e.g. Porciani \&
Givalisco 2002; for further details of our implementation see Croton
et al. 2004a).  The results of this analysis are given in Table 1.  In
Fig.~\ref{fig:sp2}, the horizontal lines show the best fit constant value for
$S_{p}$, fitted over the scales $0.71<(R/h^{-1}{\rm Mpc})<7.1$.  The
same range of scales is used to fit the $S_p$ for each order $p$.
(The choice of scales is set by the cell radii for which a reliable 
measurement of $\bar{\xi}_{6}$ is possible.)  The
dotted lines indicate the $1\sigma$ uncertainty on the fit. 
The errorbars plotted in Fig.~3 show only the diagonal component of the
full covariance matrix. The amplitudes $S_p$ are extremely correlated, 
with the principal component analysis showing that the first few 
eigenvectors contain the  bulk of the variance, indicating that 
there are typically just 2 or 3 independent points. 
Sample variance leads to measurements which could 
be coherently shifted either low or high with respect to a fixed value. 
This therefore drives the best fit value of $S_p$ 
to lie either below or above a sizeable fraction of the data points. 
For the $L_*$ sample, we note that neither $S_3$ nor $S_4$ are particularly 
well described by a constant fit (see the reduced $\chi^2$ 
values in Table 1).

For purely illustrative purposes, we have carried out the experiment
of removing the two superclusters from the $L_*$ volume limited sample and
repeating our measurement of the higher order correlation
functions. The corresponding results for the hierarchical amplitudes
are plotted using open symbols in Fig.~\ref{fig:sp2}. The upturn in
the $S_p$ values at large radii is no longer apparent. Rather than
being considered as a correction, the results of this exercise simply
serve to show the influence of the supercluster regions on our
measurements of the $\bar{\xi}_{p}$.  Where the difference matters, it
effectively indicates that the volume of even the 2dFGRS is too small
to yield a robust higher-order clustering measurement.  A further
discussion of this test is given by Croton et al. (2004a).

\begin{figure}
\plotone{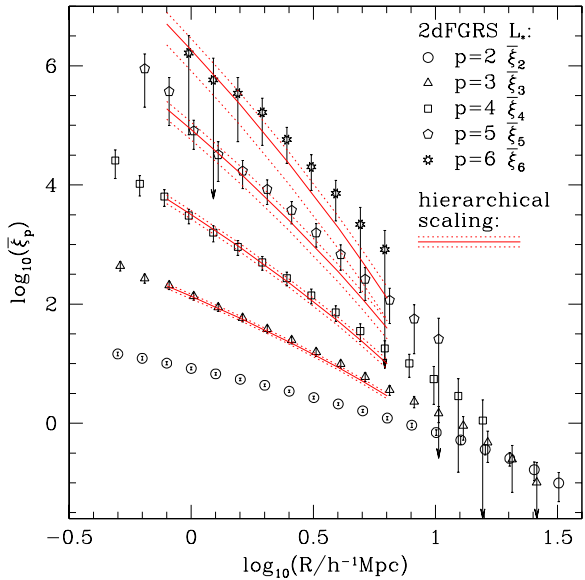}
\caption{ 
  The higher order correlation functions, $\bar{\xi}_{p}$, measured
  for $L_{*}$ galaxies in the 2dFGRS (symbols, as in Fig.~2) compared
  with the predictions of the hierarchical model (Eq. \ref{eq:sp};
  solid lines).  The hierarchical predictions are plotted only on the
  scales used to fit $S_p$. The dotted lines indicate the errors on
  these predictions, with contributions from the error on the fitted
  value of $S_p$ and on the measured variance $\bar{\xi}_{2}$.
  }
\label{fig:xip2}
\end{figure}

Armed with the best fit values of $S_p$, the hierarchical model stated
in Eq. \ref{eq:sp} can be used to make predictions for the form of the
higher order correlation functions and compare these with the measurements 
from the 2dFGRS $L_*$ galaxy sample (symbols in Fig.~\ref{fig:xip2}, 
reproduced from Fig.~\ref{fig:xip1}). The solid lines in
Fig.~\ref{fig:xip2} show the $\bar{\xi}_{p}$ predicted from the
hierarchical scaling relation (Eq. \ref{eq:sp}), assuming a constant
value for the hierarchical amplitudes, $S_p$, and using the measured
variance, $\bar{\xi}_{2}$.  The dotted lines show the uncertainty in
the theoretical predictions, derived from the 1$-\sigma$ error in the fitted
values of the $S_p$ and the error on the measured variance, $\bar{\xi}_{2}$.  
The theoretical predictions for the different orders agree spectacularly
well with the measured higher order correlation functions over the
range of scales for which the $S_p$ are fitted.

\begin{table}
  \centering
  \footnotesize
  \caption[]{The best fit values for $S_p$ and the 
    2$-\sigma$ error ($\Delta \chi^2=4$), obtained using 
    the measurements for cell radii in the range $0.71
    \le (R/h^{-1}{\rm Mpc}) \le 7.1$.  The 2$-\sigma$ errors are 
    approximately twice the size of the 1$-\sigma$ errors. 
    The final column gives the 
    reduced $\chi^2$ using the number of degrees of freedom derived from the 
    principal component analysis. 
    }
  \begin{tabular}{cccccccc} 
    \hline \hline
    {order} & {$S_p$} & $\pm 2\sigma$& $\chi^2/{\rm ndof}$\\
    \hline
     3 & 1.95 & 0.18 & 6.1 \\
     4 & 5.50 & 1.43 & 2.8 \\
     5 & 17.8 & 10.5 & 1.9 \\
     6 & 46.3 & 50   & 1.1 \\
    \hline \hline
  \end{tabular}
\end{table}

\section{Conclusions}

We have measured the higher order correlation functions of $L_*$
galaxies up to sixth order in the 2dFGRS.  Previous studies of galaxy
clustering in redshift space have been limited to fourth order
(e.g. for optically selected samples: Gazta\~{n}aga 1992; Benoist et
al. 1999; Hoyle et al. 2000: 
for infra-red selected samples: Bouchet et al. 1993; Szapudi
et al. 2000).  The volume limited sample of $L_*$ galaxies analysed
here contains 100 times more galaxies and covers 10 times the volume
of the previous best measurements from an optically selected galaxy
redshift survey (Hoyle et al. 2000).  The measured
correlation functions have a form that is in remarkably good agreement
with the predictions of hierarchical scaling, and extend to smaller
scales than those for which the perturbation theory predictions are
expected to be valid (Bernardeau et al. 2002).
A similar conclusion was reached by Croton et al. (2004b), who found 
hierarchical scaling in the reduced void probability function measured 
in the 2dFGRS.

On scales larger than about $4 h^{-1}$Mpc, there is an upturn in the 
values of $S_p$, which we have demonstrated is influenced by the 
presence of two large superclusters in the 2dFGRS (see Fig.~3).
This suggests that finite sampling affects our measurements on 
these scales.  
A similar feature was found in the angular 
Edinburgh-Durham Southern Galaxy Catalogue (EDSGC). 
Szapudi \& Gazta\~{n}aga (1998) found that 
the projected $S_p$ measured from the EDSGC 
displayed an up-turn for scales larger than $0.5$ degrees, which 
corresponds to $\approx 4 h^{-1}$Mpc at the characteristic depth 
of the survey. 
The EDSGC covers a similar part of the sky to the 2dFGRS. 
This feature in $S_p$ was not found, however, in the APM Survey, 
which covers a four times larger solid angle than the EDSGC  
(Gazta\~{n}aga 1994). 
This behaviour is not seen in the mock catalogues drawn from 
the $\Lambda$CDM Hubble Volume simulation.  Intriguingly, an 
upturn in the hierarchical amplitudes on large scales is expected 
in structure formation models with non-Gaussian initial density 
fields (Gazta\~{n}aga \& Fosalba 1998; Bernardeau et al. 2002).

Finally, we note that the variance of the distribution of cell counts for 
$L_*$ galaxies is in excellent agreement with the predictions for CDM, 
obtained from the Hubble Volume $\Lambda$CDM simulation, which includes the 
effects of peculiar motions on the clustering pattern. 
However, for cells with radii $R<10h^{-1}$Mpc the 
higher order correlation functions of $L_*$ galaxies have significantly lower 
amplitudes than the dark matter. 
This implies that the relation between the distribution of galaxies 
and the underlying dark matter may be more complicated than the popular 
linear bias model, suggesting that nonlinear contributions to the bias 
may be important on small and intermediate scales (Fry \& Gazta\~{n}aga 1993; 
see also the analyses by Conway et al. 2004 and Wild et al. 2004). 
We note that on large scales ($R>10h^{-1}$Mpc), the $\bar{\xi}_p$ measured for 
$L_*$ galaxies agree better with the $\Lambda$CDM predictions, supporting the 
conclusion reached previously, that on these scales, 
$L_*$ galaxies approximately trace the mass distribution 
(Gazta\~{n}aga \& Frieman 1994; Lahav et al. 2002; Verde et al. 2002).

We explore the distribution of galaxy counts in cells for the 
2dFGRS in more detail in Croton et al. (2004a), where we study the 
dependence of the correlation functions on luminosity.

\section*{Acknowledgements}

The 2dFGRS was undertaken using the two-degree field spectrograph on
the Anglo-Australian Telescope.
CMB is supported by a Royal Society University Research Fellowship.  
DC acknowledges a PhD fellowship from the International Max Planck 
Research School in Astrophysics.
EG acknowledges support from the Spanish Ministerio de Ciencia y
Tecnologia, project AYA2002-00850 and EC FEDER funding.  
PN acknowledges receipt of a Zwicky Fellowship.

\label{lastpage}


\begin{thebibliography}{}

\bibitem[Baugh, Gazta\~{n}aga \& Efstathiou (1995)]{1995MNRAS...274...1049}
Baugh C.M., Gazta\~{n}aga E., Efstathiou G., 1995, MNRAS, 274, 1049.

\bibitem[Benoist]{XXX}
Benoist C., Cappi A., Da Costa L.N., Maurogordato, S., Bouchet F., 
Schaeffer R., 1999, ApJ, 514, 563. 

\bibitem[Benson et al. (2001)]{XX}
Benson A.J., Frenk C.S., Baugh C.M., Cole S., Baugh Lacey C.G., 
2001, MNRAS, 327, 1041.

\bibitem[Bernardeau et al. 2002]{bcgs} 
Bernardeau F., Colombi S., Gazta\~naga E., Scoccimarro R., 
2002,  Phys. Rep., 367, 1

\bibitem[Bouchet et al. (1993)]{1993ApJ...417...36}
Bouchet F.R., Strauss M.A., Davis M., Fisher K.B., 
Yahil A., Huchra J.P., 1993, ApJ, 417, 36.

\bibitem[Colless et al.(2001)]{2001MNRAS.328.1039C} Colless M., et al.\
(the 2dFGRS Team), 2001, MNRAS, 328, 1039

\bibitem[Colless et al.(2003)]{} Colless M., et al.\
(the 2dFGRS Team), 2003, astro-ph/0306581

\bibitem[XXX]{XXX}
Conway E., et al.\ (the 2dFGRS Team), 2004, MNRAS, submitted. (astro-ph/0404276).

\bibitem[Aussies17-20England]{RWC03}
Croton D., et al.\ (the 2dFGRS Team), 2004a, MNRAS, submitted. (astro-ph/0401434).

\bibitem[XXX]{XXX}
Croton D., et al.\ (the 2dFGRS Team), 2004b, MNRAS, submitted. (astro-ph/0401406).

\bibitem[Eke et al.(2004)]{XXX} 
Eke V.R., et al.\ (the 2dFGRS Team), 2004, MNRAS, 384, 866, (astro-ph/0402567).

\bibitem[Erdogdu]{XXX}
Erdogdu P., et al.\ (the 2dFGRS Team), 2004, MNRAS, submitted. (astro-ph/0312546)

\bibitem[Evrard et al.(2002)]{XXX} 
Evrard A.E., et al.\ (the Virgo Consortium), 2002, ApJ, 573, 7.

\bibitem[Fry \& Gazta\~{n}aga(1993)]{1993ApJ...413..447} 
Fry J.~N.~\& Gazta\~{n}aga E.,\ 1993, ApJ, 413, 447

\bibitem[Gazta\~{n}aga(1992)]{1992ApJ.398..L17} 
Gazta\~{n}aga E.,\ 1992, ApJ, 398, L17

\bibitem[Gazta\~{n}aga(1994)]{1994MNRAS.268..913G} Gazta\~{n}aga E.,\ 1994, MNRAS,
268, 913

\bibitem[XXX]{XXXX} 
Gazta\~{n}aga E. \& Frieman J.,  1994, ApJ, 437, L13


\bibitem[Gazta\~{n}aga \& Fosalba (1998)]{1998MNRAS...273...L1}
Gazta\~{n}aga E., Fosalba P., 1998, MNRAS, 301, 524.

\bibitem[Hoyle, Szapudi, \& Baugh(2000)]{2000MNRAS.317L..51H} Hoyle F.,
Szapudi I., Baugh C.~M.,\ 2000, MNRAS, 317, L51

\bibitem[Juszkiewicz et al. (1993)]{1993ApJ...412...L9}
Juszkiewicz R., Bouchet F.R., Colombi S., 1993, ApJ, 412, L9.

\bibitem[Lahav, Itoh, Inagaki, \& Suto(1993)]{1993ApJ...402..387L} Lahav
O., Itoh M., Inagaki S., Suto Y.,\ 1993, ApJ, 402, 387

\bibitem[Lahav et al. (2002)]{2002MNRAS...333..961} 
Lahav O., et al. (the 2dFGRS Team), 2002, MNRAS, 333, 961.

\bibitem[Norberg et al.(2001)]{2001MNRAS.328...64N} Norberg P., et al.\
  (the 2dFGRS Team), 2001, MNRAS, 328, 64

\bibitem[Norberg et al.(2002b)]{2002MNRAS.336..907N} Norberg P., et al.\
  (the 2dFGRS Team), 2002, MNRAS, 336, 907

\bibitem[Peacock]{XXX}
Peacock J.A., et al.\ (the 2dFGRS team), 2001, Nature, 410, 169. 

\bibitem[Peebles(1980)]{1980lssu.book.....P} Peebles P.~J.~E.,\ 1980,
  The Large-Scale Structure of the Universe (Princeton Univ Press)

\bibitem[Porciani]{XXX}
Porciani C., Giavaliso, M., 2002, ApJ, 565, 24. 

\bibitem[Saunders]{Will}
Saunders W., Frenk C.S., Rowan-Robinson M., Lawrence A., Efstathiou G., 
1991, Nature, 349, 32.

\bibitem[Szapudi Gazta 1998]{XX}
Szapudi I., Gazta\~{n}aga
1998, MNRAS, 300, 493.

\bibitem[Szapudi et al. (2000)]{2000MNRAS...318...L45}
Szapudi I., Branchini E., Frenk C.S., Maddox S., Saunders W., 
2000, MNRAS, 318, L45.

\bibitem[Verde et al.(2002)]{2002MNRAS.335..432V} 
Verde L., et al.\ (the 2dFGRS Team), 2002, MNRAS, 335, 432

\bibitem[XXX]{XXX}
Wild V., et al.\ (the 2dFGRS Team), 2004, MNRAS, submitted. (astro-ph/0404275).

\end{thebibliography}
\end{document}